\documentclass[a4paper]{jpconf}
\usepackage{graphicx}
\usepackage{subfigure}

\begin{document}
\title{KLEVER: An experiment to measure \boldmath{${\rm BR}(K_L \to \pi^0 \nu \overline{\nu})$} at the CERN SPS}

\author{M.W.U.~van~Dijk on behalf of the KLEVER project}


\ead{mvandijk@cern.ch}

\begin{abstract}
The KLEVER experiment aims to measure ${\rm BR}(K_L \to \pi^0 \nu \overline{\nu})$, supplementing the ongoing NA62 measurement of ${\rm BR}(K^+ \to \pi^+ \nu \overline{\nu})$, to provide new input on CKM unitarity and potentially new physics. KLEVER is undergoing continuous development, with particular efforts focused on the design of the target and the beamline. As described here, adaptations are required relative to the K12 beamline in its current format, and a series of simulations has been performed to ensure that an adequate particle flux can be achieved while simultaneously suppressing problematic backgrounds. 

\end{abstract}

\section{Introduction}
\label{section Introduction}

\begin{figure}[!hbt]
    \center
    \includegraphics[width=0.75\linewidth]{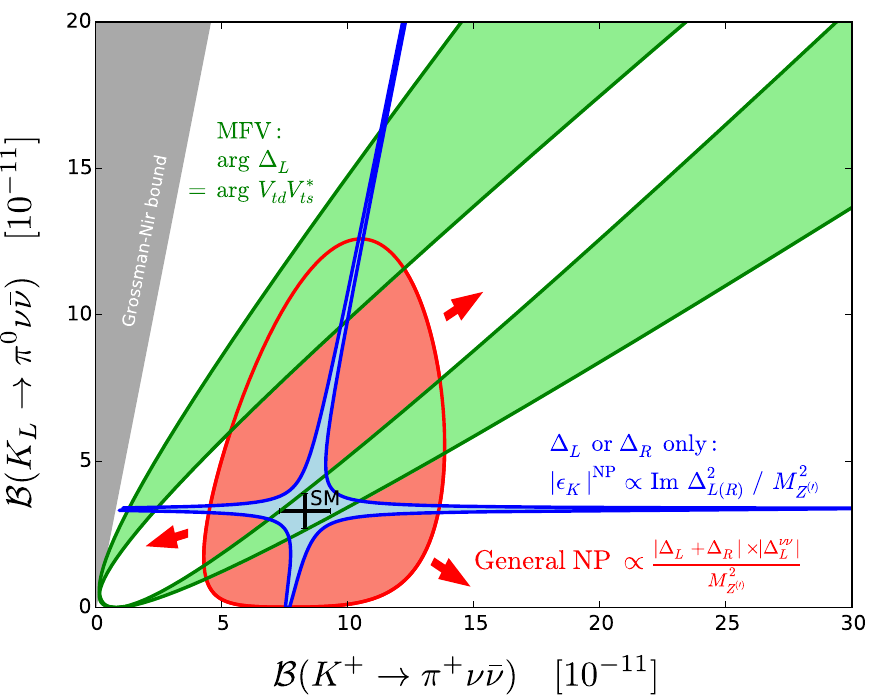}
    \caption{\label{fig_Buras_models} Standard model prediction for $BR(K \to \pi \nu \overline{\nu})$ with several potential new physics scenarios overlaid \cite{Buras_models}: CKM-like models with minimal flavour violation (green), models with flavour-violating interactions in which LH or RH couplings dominate (blue) and more general models (e.g., Randall-Sundrum, red).}
\end{figure}

A long series of both charged and neutral kaon experiments has been performed at CERN, and the natural progression of NA31 \cite{NA31_detector} and NA48 \cite{NA48_detector} has led to the current generation: the NA62 experiment \cite{NA62_detector}, measuring ${\rm BR}(K^+ \to \pi^+ \nu \overline{\nu})$. Measuring the neutral equivalent of this channel ${\rm BR}(K_L \to \pi^0 \nu \overline{\nu})$ would form the capstone in the CERN kaon program. Both are extremely rare decays for which the rates have been predicted with exacting precision in the Standard Model (SM) \cite{Buras_models}. In particular the combination of these two branching ratios has been flagged as a captivating test for new physics, being capable of determining the CKM unitarity triangle independent of inputs from $B$-physics. Current predictions in the parameter space of the two BRs are shown in Figure \ref{fig_Buras_models} (replicated from \cite{Buras_models}), with the coloured areas favoured by different classes of new physics models.

The current state-of-the-art measurement of the BR for the charged channel was obtained by BNL787/949 by means of a stopped kaon experiment, with NA62 closing in on a new measurement. The neutral channel has no measurements for the branching ratio yet; an upper limit was set by KEK391a \cite{E391a_measurement} and it is currently under investigation at the KOTO experiment ($K^0$ at Tokai \cite{KOTO_detector}). KOTO aims to observe the decay $K_L \to \pi^0 \nu \overline{\nu}$, and has recently reached a single event sensitivity of $1.3\times10^{-9}$ \cite{KOTO_measurement}. The predicted and measured values for the branching ratios are compared in Table \ref{table_BR_K}.

\begin{table}
\caption{\label{table_BR_K}Predicted and measured values of ${\rm BR}(K \to \pi \nu \overline{\nu})$}
\begin{center}
\begin{tabular}{lllll}
\br
${\rm BR}(K^+ \to \pi^+ \nu \overline{\nu})$  & Prediction    & $(8.4\pm1.0)\times10^{-11}$ & & \cite{Buras_models}\\
            & BNL787/949                    & $(17.3^{+11.5}_{-10.5})\times10^{-11}$ & & \cite{BNL_787_949_measurement}\\
            & NA62                          & $<140\times10^{-11}$  & (95\% CL) & \cite{NA62_single_event}\\
${\rm BR}(K_L \to \pi^0 \nu \overline{\nu})$  & Prediction    & $(3.4\pm0.6)\times10^{-11}$ & & \cite{Buras_models}\\
            & KEK391a	                    & $<2600\times10^{-11}$ & (90\% CL) & \cite{E391a_measurement} \\
            & KOTO                          & $<300\times10^{-11}$    & (90\% CL) & \cite{KOTO_measurement}\\
\br
\end{tabular}
\end{center}
\end{table}

The proposed site for the KLEVER experiment is the underground area ECN3 at the CERN SPS. KLEVER aims to not only observe the decay but to measure the branching ratio, and it is expected that measuring 60 events with a signal to noise ratio of 1 is achievable. To reach this goal, a total of $5\times10^{19}$ protons on target would be required, to be gathered over a period of five years, with the start of LHC Run 4 forming a potential starting point for KLEVER. In case a deviation from the SM is found, if the neutral BR is suppressed to half the expected value, this deviation could be detected with 3$\sigma$ significance in KLEVER. In the case of double the expected rate, this could be detected with 5$\sigma$ significance. \\

\section{The KLEVER experiment}
\label{section The KLEVER experiment}

The KLEVER experiment can be split into two interlinked parts: the beamline and the detectors. These have gone through an iterative design process to optimize signal acceptance and background rejection. The beam is generated by 400~GeV/c protons derived from the CERN SPS impinging on a target at an 8~mrad downward angle, generating a mixed-momentum and mixed-particle beam. The required neutral beam is selected by means of a series of collimation stations and magnetic sweepers to eject the charged components from the beam. The beam enters the volume defined by the detectors at a distance of 120~m from the target, with the final detectors placed around 240~m. The detectors are designed to handle an expected flux of primary protons on the target (POT) of $6.7\times10^{12}$ Hz, or $2\times10^{13}$ protons per spill over a 3~s spill. This is a conservative estimate relative to a more realistic spill length of 4.8~s, ensuring that fluctuations in intensity over the duration of the spill can be accounted for.

\subsection{The K12 beamline for the KLEVER experiment}

From the beam perspective, the absolute flux of $K_L$ is maximal at zero production angle. In NA48 \cite{NA48_detector}, a production angle of 2.4 mrad was used for the production of the $K_L$ beam. This particular point was chosen to balance the competing needs for high $K_L$ and low neutron fluxes. In designing the beam for KLEVER it was found that a much higher production angle, 8 mrad, was desirable, because this significantly reduces the content of $\Lambda$ in the beam. This baryon in particular forms a problematic background for KLEVER through the decay channel $\Lambda \to n \pi^0$. Choosing a larger production angle decreases $\Lambda$ production, and also softens the $\Lambda$ momentum spectrum. Because of the short lifetime of the $\Lambda$ these two effects stack and in total suppress the $\Lambda$ decays in the fiducial volume by a factor 50. Because the $\Lambda$ is production is significantly forward of $K^0$ production, when moving from 2.4 mrad to 8 mrad, the $K_L$ flux in the beam decreases less than the $\Lambda$ flux; moreover, the long lifetime of the $K_L$ means that the number of decays in the fiducial volume does not drop precipitously.

\begin{figure}[!hbt]
    \center
    \includegraphics[width=0.95\linewidth]{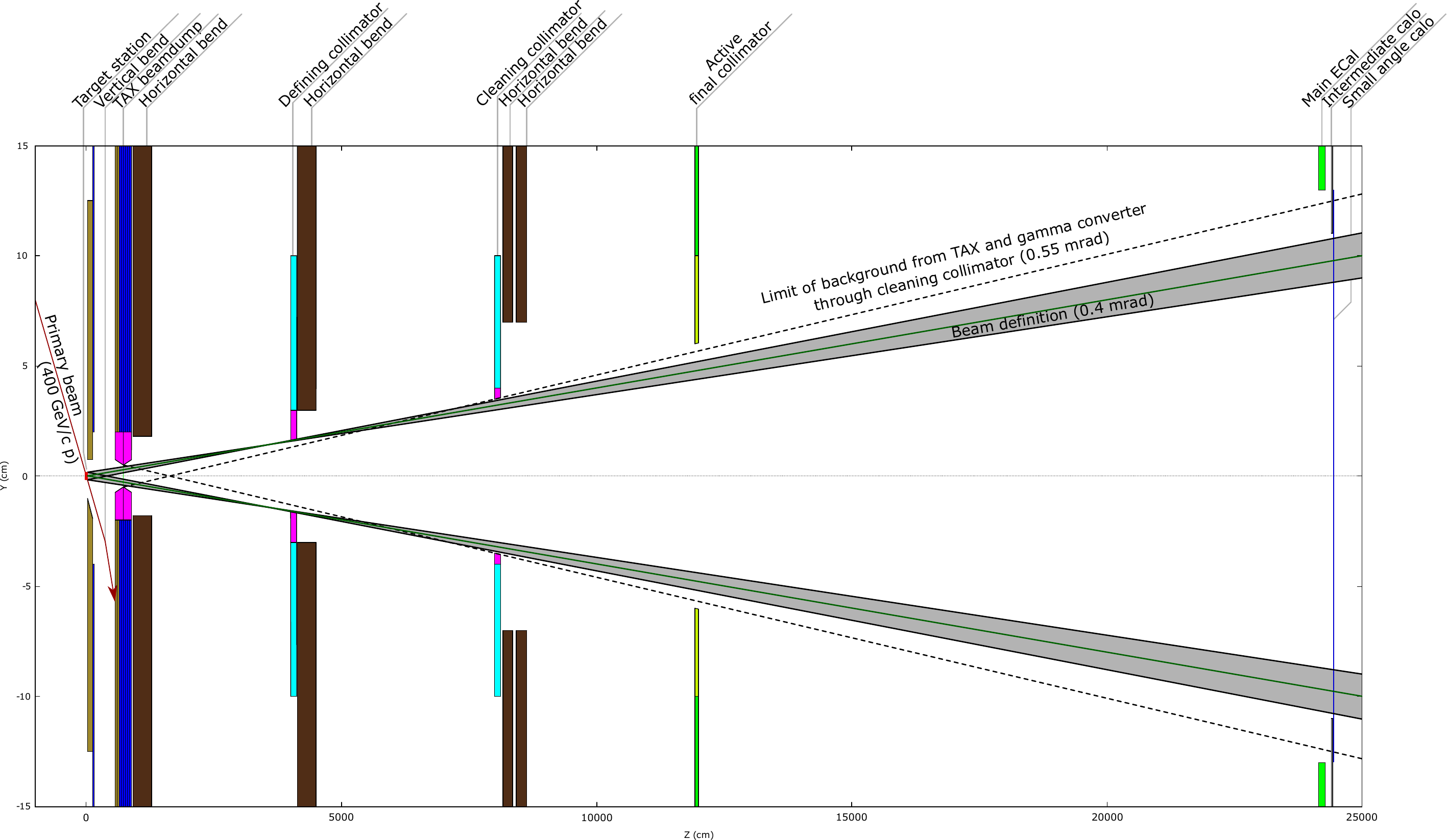}
    \caption{\label{fig_KLEVER_beamline_design} Design for the KLEVER beamline, starting from the target on the left through the four-stage collimation scheme, to the detectors on the right. Each stage of collimation except for the final (active) collimator is followed by magnetic sweeping to dump any charged particles generated.}
\end{figure}

A strong bending magnet (7.5~Tm) is placed directly after the target station to dump the primary beam with a strong downward deflection. The  mixed-purpose TAX collimator serves to dump the primary beam and unwanted secondaries and to collimate the neutral beam. A tungsten photon absorber is placed at the center of the TAX to reduce the photon component in the beam. The TAX is placed as close as possible to the target, so that only minimal numbers of muons are generated by particle decay. A second magnet after the TAX ensures that the charged particles generated on the inside faces of the collimator are swept away. The second collimator station defines the 0.4 mrad beam opening angle relative to the target, with a third collimator cleaning up particles scattered by the TAX and photon absorber that are outside the acceptable beam size. Each of these collimators is followed by a sweeping magnet to eject charged particles so that they are dumped on the next collimator downstream. The final collimator is also an active detector (AFC), so that it will detect particles interacting with its inside faces, obviating the need for magnetic sweeping. This detector defines the upstream edge of the decay volume. A schematic overview of the neutral beamline layout is given in Figure \ref{fig_KLEVER_beamline_design}.

\begin{figure}[!hbt]
    \center
    \includegraphics[width=0.5\linewidth]{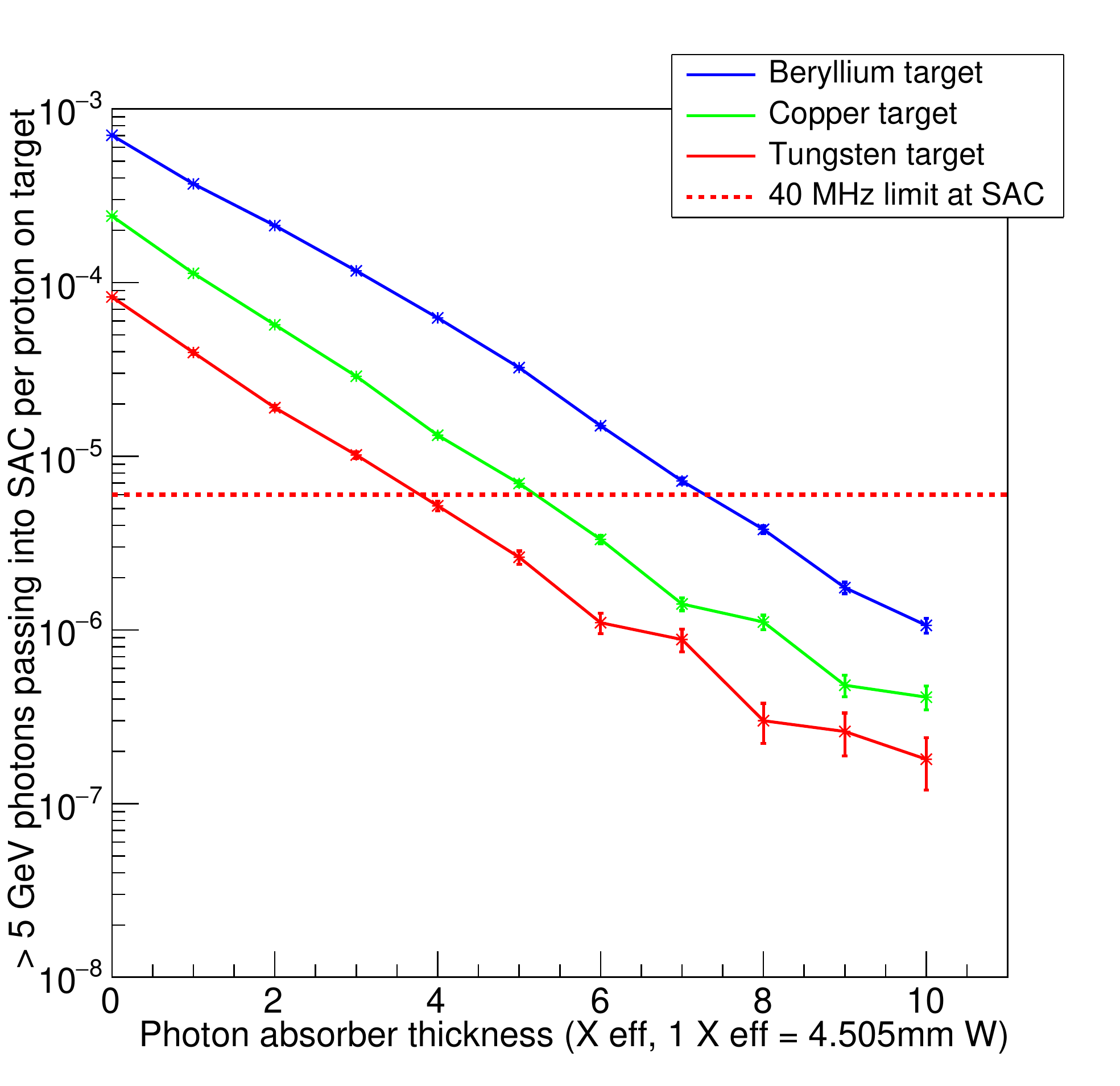}
    \caption{\label{fig_KLEVER_photon_absorber} Photon multiplicity reaching the small angle calorimeter (SAC) per POT, for a beryllium (blue), copper (green) and tungsten target (red). To reach the design criterion of $<$40 MHz of photons with E$>$5 GeV in the beam, a tungsten absorber of 32.9~mm (beryllium) / 23.3~mm (copper) / 17.2~mm (tungsten target) thickness is required. Relative to the beryllium target baseline, the $K_L$ content in the beam at the AFC can be boosted by 15\% (copper target) or 26\% (tungsten target), because fewer $K_L$s are scattered from the beam by the thinner absorber. }
\end{figure}

The tungsten absorber placed in the TAX screens high energy photons from the beam. The goal is to reduce the rate of photons in the beam with $E>5$ GeV that reach the KLEVER small angle calorimeter (see next section for description) to below 40 MHz; or below $6\times10^{-6}$ per POT. The required thickness of the absorber was determined by simulation in FLUKA \cite{FLUKA_ref1, FLUKA_ref2}. It would be possible to fabricate the KLEVER target from a different material; in particular a higher-$Z$ target would inhibit photon production by promoting pair conversion, so that a thinner photon absorber could be used. This would reduce the scattering of the $K_L$ content in the beam. The thickness of the absorber is expressed in effective absorption lengths ($X_{\rm eff}$), where 1 $X_{\rm eff} = \frac{9}{7} X_0$ since pair production is dominant in this energy regime. The photon content in the beam was studied for three target materials and is shown in Figure \ref{fig_KLEVER_photon_absorber}. 

The material of choice for the photon absorber is tungsten. Its high $Z$ gives it a rather large radiation length relative to its nuclear interaction length, making it efficient at screening photons but reasonably transparent to hadrons. It could also be fabricated with a crystalline structure. If a chosen crystal axis is aligned well with the beam axis, this would significantly improve its photon conversion properties, allowing for the absorber to be thinner. A series of beam tests has been performed at the CERN SPS in collaboration with the AXIAL project \cite{AXIAL_ref}. A tagged photon beam was used on a 10~mm tungsten crystal, and it was found that, relative to the same thickness of amorphous material, the multiplicity of charged particles was enhanced by a factor of 1.5-2.2 in the range of 10-80 GeV photon energy. This indicates increased efficiency of photon pair conversion, and a decreased radiation length. This result is preliminary; further analysis is in progress.

\subsection{The KLEVER detectors}

The KLEVER experiment consists of a series of detectors surrounding the decay volume and hermetic for all particles generated at angles of up to 100 mrad. The first detector is the upstream veto, placed 120~m downstream from the target. It screens the fiducial volume (defined from 130~m to 170~m) from particles (high energy photons in particular) generated upstream. It is constructed as a shashlik calorimeter from lead and scintillator, with the signal read out by SiPM detectors through wavelength shifting fibers. The innermost section (radius 6 to 10~cm) is referred to as the active final collimator (AFC) and is made from LYSO crystals. This high-density material functions both as a collimator to perform the final cleaning of the beam but also as an active detector, to obviate magnetic sweeping after it. The decays of $\pi^0$s to two photons inside the fiducial volume are reconstructed with the main electromagnetic calorimeter (MEC), allowing $K_L\to\pi^0\nu\bar{\nu}$ decays 
to be selected. The MEC also serves as the most important veto for events with additional photons. 

A total of 25 large angle veto (LAV) detectors are sized, placed and spaced so that all get a similar rate of detected particles. These detectors veto events with photons outside of the acceptance of the MEC, out to 100 mrad in the polar angle.  The LAVs are lead / scintillating-tile sampling calorimeters with wavelength-shifting fiber readout, based on a design for the CKM experiment \cite{Ramberg:2004en}.

\begin{figure}[!hbt]
    \center
    \includegraphics[width=0.95\linewidth]{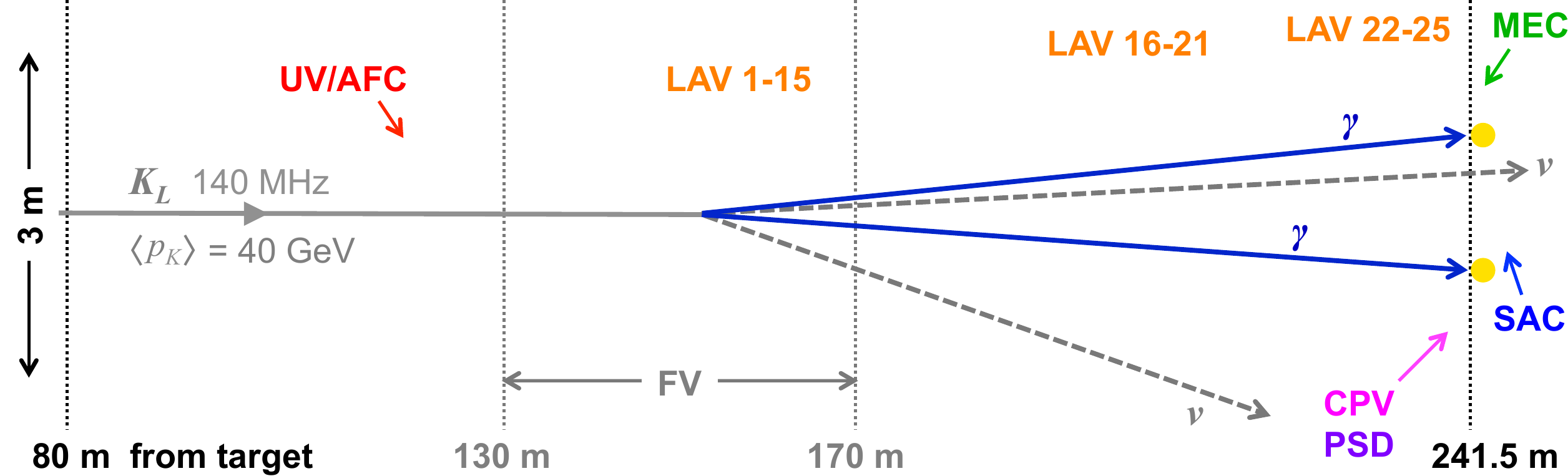}
    \caption{\label{fig_KLEVER_detectors} Layout of the KLEVER experiment, with a signal decay overlaid on top.}
\end{figure}

At the end of the detector volume, around 240~m from the target, the small angle calorimeter (SAC) is placed on the beam axis to screen it for escaping photons from $K_L$ decays, detecting particles out to a radius of 10~cm, as defined by the beam opening angle. To add further functionality, a charged particle veto (CPV), for the rejection of events with charged particles in the final state, and a pre-shower detector (PSD), to provide redundant measurements of photon trajectories, are placed just upstream of the MEC. A schematic overview of the detector is shown in Figure \ref{fig_KLEVER_detectors}. \\

The SAC has as its main design criterion that it needs to have good sensitivity to photons but be as transparent as possible to hadrons, where in particular the neutron rate ($>$400~MHz) poses difficulty. One possibility is to construct the SAC as a Si-W sampling calorimeter, potentially using thin tungsten crystals similar to those to be used for the photon absorber, promoting electromagnetic interactions and suppressing hadronic interactions.

The re-use of the NA48 krypton calorimeter (currently used in NA62) as the KLEVER MEC has been investigated, but it was found that its central passage is not sufficiently large for the expected KLEVER beam size. Additionally, it is expected that its time resolution would not be able to cope with the expected background rate, thus making a comprehensive readout upgrade mandatory. Rather, the choice was made to build the MEC from scratch, similar to the UV, as a lead / scintillator shashlik calorimeter, with the addition of so-called "spy-tiles" \cite{KLEVER_PBC}. These are thicker scintillating tiles inserted in the sampling stack, read out by separate SiPMs, and providing longitudinal information of the shower development. This information could be used for improving particle identification and time resolution of the detector. \\

\section{Signal and background}

The flux of $K_L$ in the beam sets the rate at which signal decays can be collected. Accounting for the 0.4 mrad opening angle of the beam, there are $2.1\times 10^{-5}$ $K_L$ per POT in the beam. The probability for one of these particles to decay in the fiducial volume is about 4\%, and the acceptance for a signal decay is about 5\%. At the SM branching ratio of $3.4\times10^{-11}$, with a total integrated flux of $5\times10^{19}$ POT, collecting a total of 60~SM events is expected to be an achievable target.

\begin{figure}[!hbt]
    \center
    \includegraphics[width=0.95\linewidth]{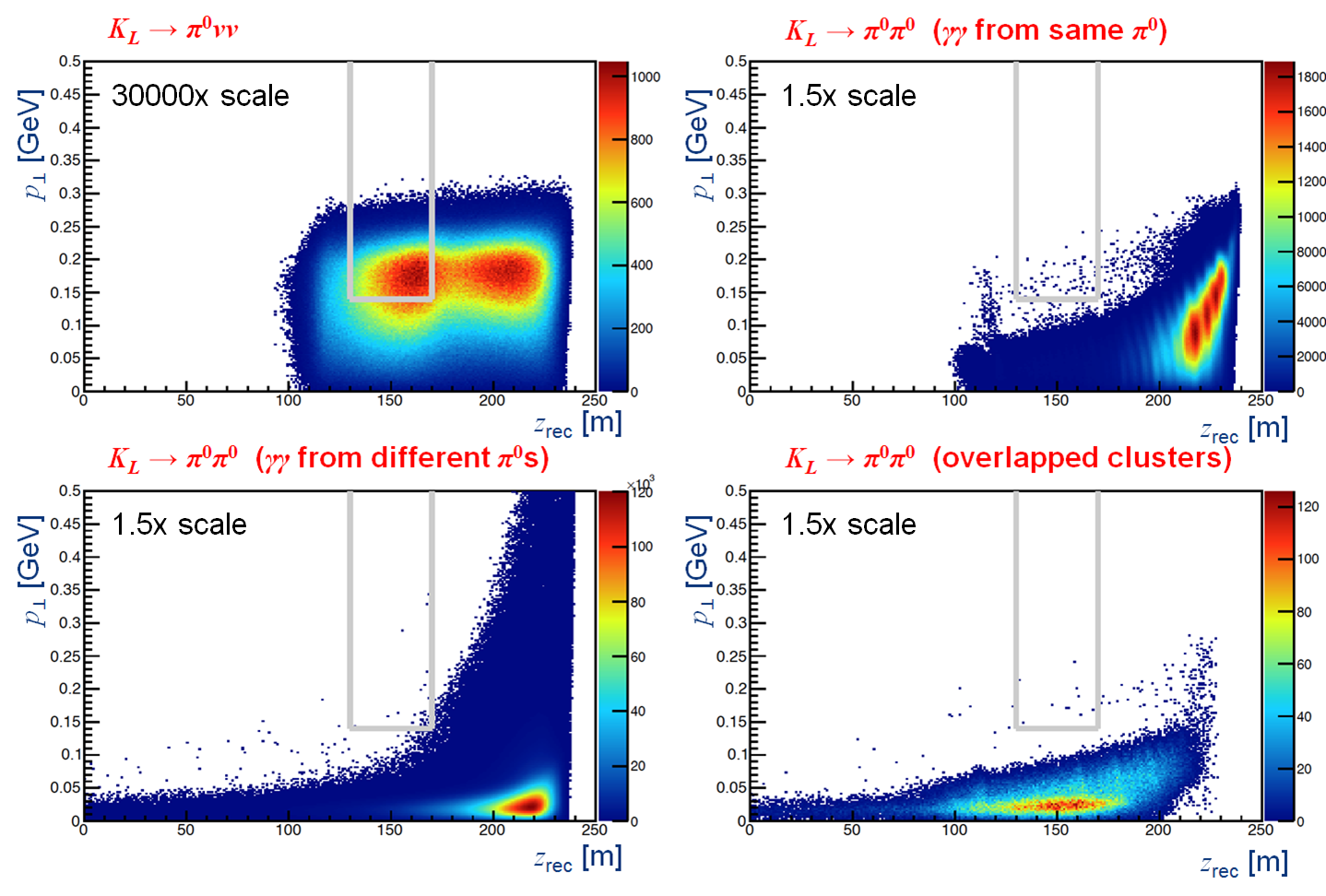}
    \caption{\label{fig_KLEVER_signal_BG} Expected signal and background from $K_L \to \pi^0\pi^0$: even, odd and overlapped clusters, shown $(z_{rec}, p_\perp)$ space, with the boundaries representing the fiducial volume and missing transverse momentum cuts overlaid. }
\end{figure}

The KLEVER signal is formed by a single reconstructed $\pi^0$ with significant missing transverse momentum. Any $K_L$ decay to charged particles would generate at least two charged particles; a reasonably efficient CPV will be able to reject these events. The decay mode $K_L \to \gamma\gamma$ is vetoed by requiring significant missing transverse momentum, while this channel would have none. With an efficient MEC, it is unlikely for four photons to be missed in the decay $K_L \to \pi^0\pi^0\pi^0$. The most problematic background is formed by the decay mode $K_L \to \pi^0\pi^0$, in the case that two of the resulting photons are missed. Three separate scenarios are identified for this background: odd (one photon is detected from each pion), even (the two photons detected are from the same pion), and overlapped clusters (two photons cannot be separated by the calorimeter and form a single cluster).
The placement of the pre-shower detector, featuring 0.5~$X_0$ of converter material and two planes of tracking ($\sigma_{x,y}~100~\mu m$), significantly reduces the odd and overlapped backgrounds.
The basic signal selection for KLEVER requires there to be no hits in the UV, AFC, LAV and SAC detectors. It also requires the decay to have occurred in the fiducial volume (determined by the reconstructed z-position of the $\pi^0$), and to have at least 140 MeV/c transverse momentum. Finally, the clusters used for $\pi^0$ reconstruction must have measured positions in the MEC at distances of more than 35~cm from the beam axis, further reinforcing the missing transverse momentum cut. Figure \ref{fig_KLEVER_signal_BG} shows the remaining signal and backgrounds for the three background classes in $(z_{rec}, p_\perp)$ space, with the boundaries representing the fiducial volume and missing transverse momentum cuts overlaid. 

\section{Conclusions and future prospects}

The KLEVER experiment is being designed as a follow-up to the NA62 experiment to measure the branching ratio of the neutral decay ${\rm BR}(K_L \to \pi^0 \nu \overline{\nu})$. This measurement by itself, but in particular in combination with the measurement of ${\rm BR}(K^+ \to \pi^+ \nu \overline{\nu})$, would constitute a powerful test of new physics. A potential layout for the beamline for KLEVER has been proposed, and an outline of the detector has been given. The design of the target is of particular interest, and the potential for significant improvements from a high-$Z$ target relative to the beryllium baseline is under investigation. Finally, a series of cuts has been proposed for extracting the signal, with further improvements expected from more sophisticated cuts and potentially from multivariate analysis. It is expected that at SM values, 60 signal events can be extracted with a 1:1 signal to background ratio.

The KLEVER experiment is expected to be sited at the CERN ECN3 cavern, taking over when NA62 has finished its data collection, currently expected to be around the end of LHC Run 3 (end of 2023). For the period up until this point, further detector development is foreseen for KLEVER. With the advent of LS3, detector construction could start in the cavern, in preparation for data taking to commence mid-2026, at the end of LS3. \\

\ack
The authors gratefully acknowledge the support of the CERN EN-EA-LE section: D.~Banerjee, J.~Bernhard, M. Brugger, N.~Charitonidis, G.L.~D'Alessandro, L.~Gatignon, A.~Gerbershagen, E.~Montbarbon, B.~Rae, M.~Rosenthal, B.M.~Veit. This work was supported by a Marie Sk\l{}odowska-Curie COFUND project of the European Commission’s Horizon 2020 Programme under contract number 665779 COFUND.


\section*{References}

\begin{thebibliography}{100}

\bibitem{NA31_detector}
Barr GD et al. (NA31 collaboration) 1993 {\it Phys. Lett.} B {\bf 317} 233-242

\bibitem{NA48_detector}
Fanti V et al. (NA48 collaboration) 2007 {\it Nucl. Instrum. Methods Phys. Res.} A {\bf 574} 433-471 

\bibitem{NA62_detector}
Cortina Gil E et al. (NA62 collaboration) 2017 {Journ. of Instr.} {\bf12} P05025

\bibitem{Buras_models}
Buras AJ, Buttazzo D, Knegjens R 2015 {\it JHEP} {\bf11} 166

\bibitem{KOTO_detector}
Tung YC (KOTO collaboration) 2016 {\it Proc. of Sci.} CD15 068

\bibitem{KOTO_measurement}
Ahn JK et al. (KOTO collaboration) 2019 {\it Phys. Rev. Let.} {\bf 122} 021802

\bibitem{BNL_787_949_measurement}
Artamonov AV et al. (E949 Collaboration) 2009 {\it Phys. Rev. } D {\bf79} 092004

\bibitem{NA62_single_event}
Ruggiero G et al. (NA62 Collaboration) 2018 {\it arXiv} 1811.08508 (to be published in PLB) 

\bibitem{E391a_measurement}
Ahn JK et al. (E391a Collaboration) 2010 {\it Phys. Rev. } D {\bf 81} 072004

\bibitem{FLUKA_ref1} 
B{\"o}hlen TT et al. 2014 {\it Nucl. Data Sheets} {\bf 120} 211-214

\bibitem{FLUKA_ref2} 
Ferrari A et al. 2005 CERN-2005-10 INFN/TC\_05/11 SLAC-R-773

\bibitem{AXIAL_ref}
Bandiera L et al. 2018 {\it Phys. Rev. Let.} {\bf 121} 021603

\bibitem{Ramberg:2004en}
  Ramberg E, Cooper P, Tschirhart R 2015 {IEEE Trans.\ Nucl.\ Sci.}  {\bf 51} 2201
  
\bibitem{KLEVER_PBC}
Ambrosino F et al. (The KLEVER project) 2019 {\it arXiv} 1901.03099

\end{thebibliography}
\end{document}